\documentstyle[12pt]{article}
\begin{document}

\rightline{UG-6/92}
\rightline{July 1992}
\vspace{2truecm}
\centerline{{\bf Realisations of $W_3$ Symmetry}}
\vspace{2truecm}
\centerline{{\bf E. Bergshoeff, H.J. Boonstra and M. de Roo}}
\bigskip
\centerline{Institute for Theoretical Physics}
\centerline{University of Groningen}
\centerline{Nijenborgh 4, 9747 AG Groningen}
\centerline{The Netherlands}
\vspace{3truecm}
\centerline{ABSTRACT}
\bigskip
We perform a systematic investigation of free-scalar realisations
of the Za\-mo\-lod\-chi\-kov $W_3$ algebra in which the
operator product of two spin-three generators contains a non-zero
operator of spin four which has vanishing norm.
This generalises earlier work
where such an operator was required to be absent. By allowing
this spin-four null
 operator we obtain several realisations of the
$W_3$ algebra both in terms of two scalars as well as in terms
of an arbitrary number $n$ of free scalars.
Our analysis is complete for the case of
two-scalar realisations.

\vfill\eject

\section{Introduction}

In recent years, there has been a lot of activity in the study
of extended conformal symmetries, better known under the name
 ``$W$-symmetries".
These symmetries constitute extensions of the Virasoro algebra
which are generically denoted by ``$W$-algebras".
$W$-symmetries can be used to clarify the structure of conformal
field theory. They also occur as a ``natural" symmetry
in a variety of physical models. Another approach is
to use $W$-symmetries for the construction of higher-spin extensions
of two-dimensional gravity (``$W$-gravity") or new string models
(``$W$-strings").

In view of the above-mentioned applications,
it is important to have a
good understanding of all possible realisations of $W$-symmetries.
The simplest example of a $W$-algebra is the $W_3$-algebra
of \cite{Zam} which contains, in addition to the spin-two Virasoro
generator $T$, a spin-three generator $W$.
Using the language of Operator Product Expansions
(OPE), the algebra is given by

\begin{eqnarray}
\label{w3}
T(z)T(w) &=& {c\over 2(z-w)^4} + {2T(w)\over (z-w)^2} +
{\partial T(w)\over z-w}+{\rm regular\ part}\,,
 \nonumber \\
T(z)W(w) &=& {3W(w)\over (z-w)^2}+{\partial W(w)\over z-w}+
{\rm regular\ part}\,,\nonumber \\
W(z)W(w) &=& {c\over 3(z-w)^6} + {2T(w)\over (z-w)^4} +
{\partial T(w)\over (z-w)^3}\\
&&+{3\over 10}
{ \partial^2 T(w)\over    (z-w)^2} + {1\over 15}
{\partial^3 T(w)\over     z-w }\nonumber\\
&&+{16\over 22+5c}\biggl ( {2\Lambda(w)\over (z-w)^2} +
{\partial\Lambda(w)\over z-w}\biggr )+{\rm regular\ part}\,.\nonumber
\end{eqnarray}
with $\Lambda = (TT) - {3\over10} \partial^2 T$.
The round brackets in $(TT)$ indicate a natural normal ordering
in terms of the Laurent modes of the generators (see, e.g.\ \cite{BBSS}).
The first equation in (\ref{w3}) gives the Virasoro algebra, while
the second equation expresses the fact that $W$ is a primary field of
spin three. The last equation tells us that the OPE of two
spin-three generators gives rise to the conformal family of
the unit operator.
The particular coefficients arising in this equation can all be fixed
by the requirement of conformal invariance. Note that in (\ref{w3})
we have used a particular normalisation of the $W$-generator, i.e.
$<WW>=c/3$, in agreement with the common convention.

In order to construct $W$-algebras and to obtain realisations of
them one can follow different strategies. One approach is to
develop a specific construction procedure, like e.g. the Miura
transformation of \cite{Fa1} or the coset
construction of \cite{Ba2}. Another approach is to start from
an Ansatz for the OPE's of a given set of abstract generators and
to require closure of the algebra (see e.g. \cite{Bl}). Alternatively,
one could start from an Ansatz for the generators of the
$W$-algebra in terms of scalar fields and then impose closure.
This has been the approach of \cite{FZ,Hu1,Rom}, where a
 systematic search for free field realisations
of the $W_3$ algebra was undertaken. In particular, starting from certain
Ans\"atze for the generators \cite{Hu1,Rom},
 the following $n$-scalar realisation was found \cite{Rom}:

\begin{eqnarray}
\label{Romans}
T &=& {1\over 2}(A_0A_0) + \sqrt 2 a_0 A_0' + T_\mu\,,\\
W &=& -{1\over 3} (A_0A_0A_0) -\sqrt {2} a_0(A_0A_0')
- {2\over 3}a_0^2A_0'' +2(A_0T_\mu) + \sqrt {2} a_0 T_\mu'\,,\nonumber
\end{eqnarray}
where $A_0$ is the derivative of a free scalar field, i.e.\
$A_0 \equiv \partial \phi_0$.
The other $n-1$ scalars are represented by $T_\mu$ which
commutes with $A_0$ and satisfies a Virasoro algebra with
central charge given by $c_\mu= {1\over4}c + {1\over2}$.
 The parameter $a_0$ is the
background charge and is related to the central charge parameter
$c$ via $c=2(1-16a_0^2)$. The resulting realisation coincides for
$n=2$ with the Fateev-Zamolodchikov (FZ) two-scalar realisation
\cite{FZ}.
It can be viewed as a natural generalisation of the FZ realisation to
an arbitrary number $n$ of scalar fields.
Note that in the definition of the nonlinear term $(TT)$ in the
$W_3$ algebra
we use a normal ordering in terms of the Laurent modes of
the generators. A normal ordering of this term
with respect to the modes of the free scalar fields was
considered in \cite{Hu3}.

The aim of this letter is to generalise the analysis of
\cite{FZ,Hu1,Rom}
by allowing  spin-four null operators in the operator product
of two spin-three generators. To be more precise, instead of the
third equation in (\ref{w3}) we require that the following OPE holds:

\begin{equation}
\label{null}
 W(z)W(w) = {\rm as\ in\ (\ref{w3}})\ + {V(w)\over (z-w)^2}
+ {{\textstyle{1\over2}}\partial V(w)\over z-w}\,,
\end{equation}
where $V$ is a spin-four null operator, i.e.\ $<VV>=0$. Of course,
strictly speaking, the algebra corresponding to (\ref{null})
is not the same as the $W_3$ algebra given in (\ref{w3}). However,
since $V$ is a null operator, it can only generate other null fields
in its OPE.
The full set of null operators constitutes an ideal
of the algebra. It is therefore consistent to set all these
null operators equal to zero and one thus obtains a
representation of the $W_3$ algebra.

Realisations
of $W$-symmetries modulo null fields
have been considered before in
the literature. For instance, they occur
in the coset construction of
\cite{Ba2} and also, in a supersymmetric context, in \cite{BCNM}.
More recently, in \cite{NQ,BK}, such realisations
were obtained, for specific values of the central charge, from a
certain nonlinear $W_\infty$ algebra \cite{BK} based upon the
coset $SL(2,R)/U(1)$. This algebra is related to the
parafermion current algebra of \cite{Za2}.
{}From a somewhat different
point of view, extensions of the $W_3$ algebra with null generators
 have  occurred
recently in a study of certain singular contractions of $W$-algebras
\cite{HP}.

In our analysis of the $W_3$ algebra,
 we have restricted ourselves in the following two ways.
First of all, we only consider
spin-four null operators. In principle, one could also allow for
spin-two null operators in the OPE of two spin-three generators.
However, since in most formulations of
$W$-algebras every spin occurs only once, it is less natural
to allow for spin-two operators in addition to the Virasoro generator.
Secondly, we only consider free field realisations.
 We will not consider the inclusion of vertex operators
 in the Ansatz as was done in \cite{SSY}.

\bigskip
\section{Ans\"atze}

Our starting point is the following
free field Ansatz for the spin-two and spin-three generators
of the $W_3$-algebra \cite{Hu1,Rom}:
\begin{eqnarray}
T&=&{1\over2}g_{ij}(A^i A^j)+\sqrt{2}a_i {A^i}'\,,\label{2.1.a}\\
 W&=&{1\over3}d_{ijk}(A^i A^j A^k)+2\sqrt{2}e_{ij}(A^i {A^j}')+
 2f_i {A^i}''
\,,\label{2.1.b}
\end{eqnarray}
where $A^i\equiv\partial\phi^i$ and the $\phi^i\ (i=0,\dots,n-1)$
 represent a set of $n$ free scalar fields and $g_{ij}, a_i,
d_{ijk}, e_{ij}$ and $f_i$ are yet undetermined coefficients.
 The $A^i$ satisfy the OPE
\begin{equation}
A^i (z)A^j (w)={g^{ij}\over(z-w)^2}+{\rm regular\
 part}\,,
\end{equation}
where $g^{ij}$ is the inverse of $g_{ij}$.
 Our conventions are slightly different from those of \cite{Rom}.
 Note that with the above Ansatz the spin-two
 generator $T(z)$ already satisfies the Virasoro
 algebra with central charge
 $c=n-24a_i a^i$.

Following \cite{Rom}, we split
the $n$-component index $i$ into ``0'' and
 an $(n-1)$-component index $\mu$ and take the coefficients $d_{ijk}$
to be
\begin{equation}
d_{000}=s\ ,\ \ d_{0\mu\nu}=-sg_{\mu\nu}\,,\label{2.3}
\end{equation}
where the parameter $s$ is fixed by the choice of normalisation of
the $W$ generator.
The expression for the $d$-coefficients is a solution to
\begin{equation}
\label{classical}
d_{(ij}^{\ \ \ m} d_{kl)m}=s^2 g_{(ij}g_{kl)}\,.
\end{equation}
The latter equation guarantees the closure of the classical
 version $w_3$ of the
 $W_3$ algebra \cite{Hu1}. In the analysis of
 \cite{Rom}, equations for the unknown coefficients
 in (\ref{2.1.a}) and (\ref{2.1.b}) were found by demanding
 that the generators satisfy the  $W_3$ algebra given in (\ref{w3}),
i.e.\ without spin-four null operators.
It was subsequently shown that these equations were solved by the
$n$-scalar realisation given in (\ref{Romans}).

We now consider the same Ansatz (\ref{2.1.a},\,\ref{2.1.b}), but
instead require that the generators satisfy the $W_3$ algebra
modulo a spin-four null operator as indicated in (\ref{null}).
This allows us to take the following less
restrictive Ansatz for the coefficients $d_{ijk}$:
\begin{equation}
d_{000}=s\ ,\ \ d_{0\mu\nu}=tg_{\mu\nu}\,,\label{2.6}
\end{equation}
with $s$ and $t$ free parameters (although one of them may be fixed
by choosing a normalisation for $W$).

The following three equations
 have to be satisfied in order that $W$ is primary
 w.r.t. $T$:
\begin{eqnarray}
d^j_{\ ji}-24e_{ij}a^j+12f_i=0\,,\label{2.7.a}\\
2e_{(ij)}-d_{ijk}a^k=0\,,\label{2.7.b}\\
3f_i-2a^j e_{ji}=0\,.\label{2.7.c}
\end{eqnarray}
For more details, see \cite{Rom}. On the fourth order pole of the
 OPE of $W$ with itself a primary spin-two operator shows up
 besides a multiple of the energy momentum tensor. We require that
this
 operator vanishes because we want $T$ to be the only spin-two
 operator in the algebra. This leads to the following equation
\begin{equation}
d_i^{\ kl}d_{jkl}+12d_{ijk}f^k-24e_i^{\ k}e_{jk}=
{3\over2c}N_3 g_{ij}
\,.\label{2.8}
\end{equation}
In (\ref{2.8}) $N_3$ is the norm of the operator $W$,
which we prefer not to fix for the moment:
\begin{equation}
N_3\equiv <WW>={2\over3}\Bigl[ d_{ijk}d^{ijk}-72e_{ij}e^{ij}
-48e_{ij}e^{ji}+720f_i f^i\Bigr]\,.\nonumber
\end{equation}
In \cite{Rom}
 two more equations were used, which guaranteed the vanishing of a
 primary spin-four operator $V$ in the OPE of $\,W(z)W(w)$. Instead, we
 will allow such a spin-four operator, but
 only if it is null. This requirement leads to
 one more, rather complicated, equation which we have given in Appendix
A. We will refer to this equation as the spin-four equation.

We conclude that the full set of equations that has to be satisfied by
the Ansatz (\ref{2.1.a}),\,(\ref{2.1.b}) and (\ref{2.6}) is given by
 equations (\ref{2.7.a}-\ref{2.8}) and the spin-four equation
 which can be found in Appendix A. The general analysis of these
equations is rather complicated. Among the solutions one should of
course find, as a special case, those of \cite{Rom}
 which are characterized
by taking $s=-t$ in  (\ref{2.6}) and $V\equiv 0$, i.e.\ no spin-four null
operator. We will now discuss the new solutions we obtained.

\bigskip

\section{Solutions}

Our strategy is to first solve
equations (\ref{2.7.a}-\ref{2.8}) and afterwards impose the
 the spin-four
 equation. It is convenient to distinguish between the two
 cases corresponding to $a_0\neq 0$ and $a_0=0$.  From now on we will
take $t=1$ as a choice of normalisation. Note that in general this
differs from the standard normalisation $<WW>=c/3$.
 For $a_0\neq
 0$ (case I) we find:
\begin{eqnarray}
\hspace{-2.4cm}I\hspace{3cm}e_{00}={\textstyle {1\over2}}sa_0\,,
\ \ \ e_{\mu 0}=0\,,\ \ \ e_{0\mu}=a_\mu\,,\nonumber\\
e_{(\mu\nu)}={\textstyle {1\over2}}a_0g_{\mu\nu}\,,
\ \ \ e_{[\mu\nu]}=0\,,
\nonumber\\
f_0={\textstyle {1\over3}}sa_0^2\,,\ \ \ f_\mu=a_0 a_\mu\,.
\end{eqnarray}
Besides the Romans solution, corresponding to $s=-1$, these
equations have the following other solutions as well:
\begin{eqnarray}
\hspace{-2cm}&I&\hspace{20pt}
a_0^2={s-2\over2(s-3)}\,,\ \ \ a_\mu a^\mu=
{-3s^2+4s+3+n(s-3)\over24(s-3)}\,,\\
\hspace{-2cm}& &\hspace{110pt}c=3s-7\,,\label{3.3.b}
\end{eqnarray}
where the parameter $s$ is still undetermined. We
now substitute these solutions into the spin-four equation.
It turns out that this equation is satisfied
 only for the values $s=7/3,\ 5/3,\ -1, \ 5/2$ and $13/5$.
 For $s=7/3$ and $s=5/2$,
 corresponding to $c=0$ and $c=1/2$,
 respectively, $W$ turns out to be a null field as well, and we
 will not consider these cases further. For $s=-1$ we get
 the Romans solution for $c=-10$. The two new solutions we find
 are given
 by $s=5/3\ (c=-2)$ and $s=13/5\ (c=4/5)$.
In appendix B the
explicit form of these realisations is given for $n=2$.

We note that
the $c=4/5$ realisation has an imaginary background charge $a_0$.
 In order to obtain real
 coefficients in the realisation it is necessary to perform the
 redefinitions $A_0\rightarrow iA_0$ and $W\rightarrow iW$.
 The result is
 a ``non-compact'' realisation where the quadratic $A_0$ part
 in $T$ has a minus sign.

 A general feature of the case I solutions is that
 the $W_3$ generators take on the form
\begin{eqnarray}
\hspace{-1cm}T&=&{1\over2}(A_0 A_0)+
\sqrt{2}a_0 A_0'+T_\mu\,,\label{3.4.a}\\
\hspace{-1cm}W&=&{1\over3}s(A_0 A_0 A_0)+\sqrt{2}sa_0(A_0 A_0')+
{2\over3}sa_0^2 A_0''
+2(A_0 T_\mu)+\sqrt{2}a_0 T_\mu'\,,\label{3.4.b}
\end{eqnarray}
where $T_\mu$ is the energy momentum tensor corresponding to
 the $n-1$ fields $A_\mu$ with central charge
\begin{equation}
c_\mu =-s(1-8a_0^2)\,.\label{3.5}
\end{equation}
The total central charge is
\begin{equation}
c=c_0+c_\mu=1-24a_0^2-s(1-8a_0^2)\,.\label{3.6}
\end{equation}
So there is one scalar that appears explicitly
 in (\ref{3.4.a},\,\ref{3.4.b}), and the rest enters only via their
 energy momentum tensor $T_\mu$.
This situation also occurs in the Romans solution (\ref{Romans}).
 We note that
 for both the Romans solution (\ref{Romans})
 at $c=-2$  as
well as the case I $c=-2$ solution given in (\ref{3.4.a},\,\ref{3.4.b}),
$T_\mu$ is null and the $A_0$ part
 becomes the one scalar realisation of $W_3$ \cite{BCNM}.

We next consider solutions of eqs.\ (\ref{2.7.a}-\ref{2.8})
 for $a_0=0$ (case II). From equations (\ref{2.7.a}-\ref{2.8}) we
deduce that
\begin{eqnarray}
\hspace{-3cm}II\hspace{120pt}
e_{00}=0\,,\ \ \ e_{0\mu}+e_{\mu 0}=a_\mu\,,
\nonumber\\
e_{0\mu}a^\mu={\textstyle
{1\over4}}a_\mu a^\mu+{\textstyle {1\over32}}
(s+n-1)\,,\nonumber\\
e_{(\mu\nu)}=0\,,\ \ \ e_{[\mu\nu]}a^\nu=0\,,\ \ \
e_{[\mu\nu]}e_0^{\ \nu}=0\,,\nonumber\\
f_0={\textstyle {1\over2}}
a_\mu a^\mu-{\textstyle {1\over48}}(s+n-1)\,,
\ \ \ f_\mu=0\,.\label{3.7}
\end{eqnarray}
Furthermore, the background charges and the central charge
are given by
\begin{eqnarray}
\hspace{-2.8cm}&II&\hspace{78pt}
 a_\mu a^\mu={\textstyle {1\over 24}} (n-3s+7)\,,
\ \ \ c=3s-7\,.\label{3.9}
\end{eqnarray}
We also obtain expressions for the contractions $e_{0\mu}e^{0\mu}$
 and $e_{[\mu\nu]}e^{[\mu\nu]}$. Since they are rather involved we
will not give them here.
We still have to impose the spin-four equation.
  We were able to simplify this equation only for $n=2$
and have not analysed it for general values of $n$.
 For $n=2$ the spin-four equation becomes, rewritten in
 terms of $c$ using (\ref{3.9}):
\begin{equation}
<VV>={16c(2+c)(7+c)(10+c)^2(-{1\over2}+c)(-4+5c)\over
27(-2+c)^2(22+5c)}=0\,.\label{3.10}
\end{equation}
{}From the series of roots of (\ref{3.10}) the values $c=0,\ -7,\
 1/2$ make $W$ a null field as well, and for $c=-10\ (s=-1)$ we get
 a FZ realisation. The new solutions occur
 again for $c=-2$ and $c=4/5$.
 The case II $c=-2$ and $c=4/5$
 realisations also appear in \cite{BK} as specific truncations of a
 non-linear $W_\infty$ algebra.
They can also be derived from the second realisation mentioned in
a footnote of the paper by Fateev and Zamolodchikov \cite{FZ}.
 The explicit form of the solutions can be found in Appendix B.

Unlike the case I realisations the case II
 realisations are not of the form (\ref{3.4.a},\,\ref{3.4.b}), i.e.
 there exists no $SO(2)$ redefinition of the fields such
 that (\ref{3.4.a},\,\ref{3.4.b}) is obtained. It is therefore not clear
whether these solutions can be generalised to $n\ge 2$ scalars.

\section{Generalisations}

We now discuss generalisations of the case I and case II realisations.
First, consider the $c=-2$ one-scalar
realisation of \cite{BCNM}:
\begin{eqnarray}
T_0 &=& {1\over2}(A_0 A_0)+{1\over2}A_0'\,,\\
W_0 &=&
-{2\over3}(A_0 T_0)-{1\over6}
T_0'\,.
\end{eqnarray}
We now add an extra
 energy momentum tensor, denoted by ${\tilde T}$, to the above system
 that commutes with $A_0$ and which is null.
 We then make the following Ansatz
 for $W$:
\begin{eqnarray}
T&=&T_0+\tilde{T}\,,\\
W&=&W_0+d_1(A_0\tilde{T})+d_2\tilde{T}'\,.\label{Wdd}
\end{eqnarray}
Since $\tilde{T}$ commutes with $T_0$ the total central charge is
 given by $c=-2$. The requirement
 that $W$ is primary w.r.t. $T$ can be shown to imply $d_1=4d_2$.
 Next, in order to get rid of a primary spin-two field in the
 OPE $\,W(z)W(w)$, which occurs in addition to $T$,
 the following quadratic equation has to be satisfied:
\begin{equation}
20d_2^2-4d_2-3=0
\end{equation} with roots $1/2$ and $-3/10$.
If we represent $\tilde{T}$ in terms
 of $n-1$ scalar fields ($n\geq2$) then we obtain for $d_2=1/2$
 the Romans realisation at $c=-2$ and for $d_2=-3/10$ the case I
 $c=-2$ realisation (cf.\ (\ref{3.4.b})).
Note that if, in the above example, we do not modify $W_0$
(i.e. $d_1=d_2=0$ in (\ref{Wdd})), the algebra also closes modulo
null operators. However, in this case also a spin-two null
operator is present in $W(z)W(w)$.

We now perform the same procedure
starting from the Romans realisation (\ref{Romans}) for
 arbitrary $c$.
Again we add a null energy momentum tensor ${\tilde T}$ to the generators
in such a way that they remain primary. We thus obtain
\begin{eqnarray}
\hspace{-1cm}T&=&T_0+T_\mu+\tilde{T}\,,\label{4.4.a}\\
\hspace{-1cm}W&=&W_0+2(A_0 T_\mu)+
\sqrt{2}a_0T_\mu'+d\bigl[(A_0\tilde{T})+
{1\over2}\sqrt{2}a_0
\tilde{T}'\bigr]\,,\label{4.4.b}\\
\hspace{-1cm}T_0&=&{1\over2}(A_0 A_0)
+\sqrt{2}a_0 A_0'\,,\label{4.4.c}\\
\hspace{-1cm}W_0&=&
-{1\over3}\Bigl[2(A_0 T_0)+\sqrt{2}a_0 T_0'\Bigr]\,.\label{4.4.d}
\end{eqnarray}
 The central charges are given by (take $s=-1$ in (\ref{3.5})
 and (\ref{3.6}))
\begin{eqnarray}
c_0&=&1-24a_0^2={\textstyle {3\over4}}c-{\textstyle {1\over2}}
\,,\label{4.5.a}\\
c_\mu&=&1-8a_0^2={\textstyle
{1\over4}}c+{\textstyle {1\over2}}\,.\label{4.5.b}
\end{eqnarray}
 The requirement that the additional spin-two primary field that
 occurs in $WW$ vanishes, now leads to the following equation
\begin{equation}
-2+{\textstyle{1\over2}}d^2+a_0^2
(-{\textstyle{3\over2}}d^2-2d+10)=0\,,
\end{equation}
with roots
\begin{equation}
d=2\,,\hspace{1cm} d=-2{1-5a_0^2\over 1-3a_0^2}\,.
\end{equation}
For the $d=2$ solution,
$\tilde{T}$ can be absorbed into $T_\mu$ and we obtain
the Romans realisation.
 The second solution for $d$ does
not fit within the Ansatz (\ref{2.6}), which
is why we did not find this solution before.
 For $c=-2$, $a_0^2=1/8$, $T_\mu$ is null
 and can be consistently put to zero, and we find
that for this value of $c$ the second solution
reduces to the case I $c=-2$ realisation.

In principle, one could generalise the case II realisations from $n=2$
to arbitrary $n$ by the same procedure. One adds a null
 field $\tilde{T}$ to $T_0$ and
 adds $d_1(A_0\tilde{T})+d_2(A_1\tilde{T})+d_3\tilde{T}'$ to $W_0$.
 Making $W$ primary fixes one parameter, and the spin-two absence
 implies a quadratic equation in the two remaining
 parameters. We have not attempted to investigate systematically
 the solutions to this equation.

\bigskip

\section{Comments}

We have performed a systematic investigation of free-field
realisations of the
$W_3$ algebra where we allow in the OPE of two spin-three
generators a spin-four null field. Our starting point was
a free field Ansatz for the generators. Closure of the algebra then
led to a set of equations for the coefficients occurring in the Ansatz.
We analyzed these equations and gave several solutions to them.
Besides the Romans solution (see (\ref{Romans})),
we found further two-scalar solutions (case II) as well as n-scalar
solutions (case I and the second solution of
eqs. (\ref{4.4.a}-\ref{4.4.d})).

Since we used a specific Ansatz, our analysis is not exhaustive.
Only in the case of two-scalar realisations were we able to verify that
our analysis
is complete. Besides the FZ realisation we found four more realisations
whose explicit form can be found in Appendix B. Two of these solutions
also occur in the work of \cite{FZ,NQ,BK}. It would be interesting
to see whether the other two solutions could be understood
from other construction procedures as well.

Finally, one could consider the classical limit of our results.
In the case of the Romans
realisation one obtains in this limit a realisation
of a classical version $w_3$ of the $W_3$ algebra.
This is consistent with the fact that the Ansatz of \cite{Hu1,Rom}
satifies the identity
(\ref{classical}) which guarantees the closure of the classical
$w_3$ algebra \cite{Hu1}.
Our Ansatz does not satisfy (\ref{classical}) and
therefore, to obtain closure in the classical limit, one should
include the whole ideal of null operators generated by the
spin-four operator $V$.
For the case II solutions, this leads
to the classical limit of the nonlinear $W_\infty$
algebra of \cite{BK}.
It would be interesting to see
which classical algebras the case I realisations lead to.

\bigskip
\bigskip
\bigskip
\centerline{{\bf ACKNOWLEDGEMENTS}}

\bigskip

We would like to thank Kris Thielemans for explaining to us how
to use his
Mathematica package for computing  operator product expansions
 \cite{KT}.
One of us (E.B.) would like to thank Adel Bilal,
Peter Bouwknegt, Bernard de Wit, Alexander
Sevrin and Shawn Shen for useful discussions and the CERN
Theory Division for its hospitality during a visit in July.
The work of H.J.B. was performed as part of the research program
of the ``Stichting voor Fundamenteel Onderzoek der Materie'' (FOM).
The work of E.B. has been made possible by a
fellowship of the Royal Netherlands Academy of Arts and Sciences
(KNAW).

\appendix

\section{The Spin-Four Equation}

To determine the spin-four equation mentioned in section 2,
we must first calculate the
expression for the spin-four operator $V$. This expression can be found
from eq.\ (3.12) in \cite{Rom} by subtracting the descendents of
the energy-momentum tensor. Next, it is a straightforward exercise
to calculate the norm of $V$ and require it to be zero.
We thus find the following spin-four equation:

\begin{eqnarray*}
<VV>&=&24S^{ijkl}S_{ijkl}+30S_i^{\ ikl}S_{j\ kl}^{\ j}
-280S_i^{\ ikl}S_{j\ k}^{\ j\ m}a_l a_m\\
& &-60\sqrt{2}S_i^{\ ikl}T_{klm}a^m+24\sqrt{2}
S_i^{\ ikl}T_{mkl}a^m\\
& &+{560\over3}\sqrt{2}S_i^{\ ikl}
(T_{mnl}+2T_{lmn})a_k a^m a^n
-12T^{ijk}T_{ijk}+\\
& &-16T^{ijk}T_{ikj}+60T_{ijk}
T^{ij}_{\ \ l}a^k a^l-48T_{ijk}T_l^{\ ij}
a^k a^l\\
& &+{328\over3}T_{kij}T_l^{\ ij}a^k
a^l+{104\over3}T_{kij}T_l^{\ ji}a^k a^l\\
& &-{560\over9}(T_{ijm}T_{kl}^{\ \ m}+
4T_{imj}T_{kl}^{\ \ m}+4T_{imj}T_{k\ l}
^{\ m})a^i a^j a^k a^l \\
 &=& 0\,,
\end{eqnarray*}
where $S$ and $T$ are given by
\begin{eqnarray*}
S_{ijkl}&=&d_{(ij}^{\ \ \ m}
d_{kl)m}-{24N_3\over c(22+5c)}  g_{(ij}g_{kl)}\,,
\\
T_{ijk}&=&4\sqrt{2}\Bigl(-2d_{ij}^{\ \ l}
e_{[kl]}+2e_{(i}^{\ \ l}d_{j)kl}
-{24N_3\over c(22+5c)}  g_{ij}a_k \Bigr)\,.
\end{eqnarray*}

\section{Two-scalar realisations}

For the case of two scalars $(n=2)$ we find
all possible
 realisations of $W_3$ that close modulo a
 non-zero spin-four null field.
We find four different realisations. Firstly, the case I $c=-2$
realisation is given by
\begin{eqnarray}
T&=&{1\over2}(A_0 A_0)+{1\over2}(A_1 A_1)+{1\over2}A_0'
+{1\over6}\sqrt{3}A_1'
\,,\nonumber\\
W&=&{5\over9}(A_0 A_0 A_0)+{5\over6}
(A_0 A_0')+{5\over36}A_0''\nonumber\\
& &+(A_0 A_1 A_1)+{1\over3}
\sqrt{3}(A_0 A_1')+{1\over2}(A_1 A_1')+
{1\over12}\sqrt{3}A_1''\,.\nonumber
\end{eqnarray}
Secondly, the case I
$c=4/5$ realisation in a real basis is given by
\begin{eqnarray}
T&=&-{1\over2}(A_0 A_0)+{1\over2}
(A_1 A_1)+{1\over2}\sqrt{6}A_0'+{2\over5}
\sqrt{10}A_1'\,,\nonumber\\
W&=&{13\over15}(A_0 A_0 A_0)-
{13\over10}\sqrt{6}(A_0 A_0')+{13\over10}A_0''
\nonumber\\
& &-(A_0 A_1 A_1)-{4\over5}\sqrt{10}
(A_0 A_1')+{1\over2}\sqrt{6}(A_1 A_1')+
{1\over5}\sqrt{60}A_1''\,.
\nonumber
\end{eqnarray}
The above realisations are obtained
 from (\ref{3.4.a},\,\ref{3.4.b}) by substituting the appropriate
 values for the parameters and by realising $T_\mu$ in terms of
$A_1$, the derivative of the scalar field $\phi_1$.

Next, the case II $c=-2$ realisation is given by:
\begin{eqnarray}
T&=&{1\over2}(A_0 A_0)+{1\over2}(A_1 A_1)+
{1\over 3}\sqrt {3} A_1'\,,\nonumber\\
W&=&{5\over9}(A_0 A_0 A_0)+
(A_0 A_1 A_1)+{1\over 2}\sqrt {3}(A_0 A_1')
\nonumber\\
& &+{1\over6}\sqrt {3}(A_0' A_1)
+{1\over18}A_0''\,.\nonumber
\end{eqnarray}
Finally, the case II $c=4/5$ realisation is given by
\begin{eqnarray}
T&=&{1\over2}(A_0 A_0)+{1\over2}(A_1 A_1)+
{1\over 10}\sqrt {10} A_1'\,,\nonumber\\
W&=&{13\over 15}(A_0 A_0 A_0)+
(A_0 A_1 A_1)+{1\over2}\sqrt {10}(A_0 A_1')
\nonumber\\
& &-{3\over 10}\sqrt {10}(A_0' A_1) - {1\over10}A_0''\,.\nonumber
\end{eqnarray}
For the $n=2$, $c=-2$ realisations $<WW>=-{25\over9}$. The
$c=4/5$ realisations have $<WW>={52\over75}$ (case I) and
$<WW>=-{52\over75}$ (case II).

\hspace{3cm}

\end{document}